\pdfoutput=1
\documentclass{article}

\usepackage{fancyhdr}
\makeatletter
\fancypagestyle{firstpage}{
  \fancyhf{}

  \fancyfoot[C]{\parbox{\textwidth}{\centering\scriptsize
  © 2026 IEEE. Personal use of this material is permitted. Permission from IEEE must be obtained for all other uses, in any current or future media, including reprinting/republishing this material for advertising or promotional purposes, creating new collective works, for resale or redistribution to servers or lists, or reuse of any copyrighted component of this work in other works.}}
}
\makeatother

\usepackage{spconf}            
\usepackage[T1]{fontenc}
\usepackage[utf8]{inputenc}    
\usepackage{microtype}
\usepackage{amsmath}
\usepackage{amssymb}
\usepackage{graphicx}
\usepackage{booktabs}
\usepackage{multirow}
\usepackage{caption}
\usepackage{subcaption}
\usepackage{xspace}
\usepackage{url}
\usepackage{color}
\usepackage{colortbl}
\usepackage{enumitem}
\usepackage{float}
\usepackage{placeins}
\usepackage{stfloats}
\usepackage{verbatim}
\usepackage{textcomp}          
\usepackage[normalem]{ulem}
\useunder{\uline}{\ul}{}

\usepackage[hidelinks]{hyperref}

\title{Wave-GMS: Lightweight Multi-Scale Generative Model for Medical Image Segmentation}

\name{Talha Ahmed, Nehal Ahmed Shaikh, Hassan Mohy-ud-Din\thanks{This work was supported by a grant from the Higher Education Commission of Pakistan as part of the National Center for Big Data and Cloud Computing. Email: \href{mailto:hassan.mohyuddin@lums.edu.pk}{hassan.mohyuddin@lums.edu.pk}}}
\address{School of Science and Engineering, Lahore University of Management Sciences, Lahore, Pakistan}

\begin{document}
\thispagestyle{firstpage}
\topmargin=0mm 
\ninept

\maketitle

\begin{abstract}
For equitable deployment of AI tools in hospitals and healthcare facilities, we need Deep Segmentation Networks that offer high performance and can be trained on cost-effective GPUs with limited memory and large batch sizes. In this work, we propose Wave-GMS, a lightweight and efficient multi-scale generative model for medical image segmentation. Wave-GMS has a substantially smaller number of trainable parameters, does not require loading memory-intensive pretrained vision foundation models, and supports training with large batch sizes on GPUs with limited memory. We conducted extensive experiments on four publicly available datasets (BUS, BUSI, Kvasir-Instrument, and HAM10000), demonstrating that Wave-GMS achieves state-of-the-art segmentation performance with superior cross-domain generalizability, while requiring only $\sim$2.6M trainable parameters. Code is available at \url{https://github.com/ATPLab-LUMS/Wave-GMS}. \\
\end{abstract}        
\vspace{-0.2cm}
\section{Introduction}
\label{sec:intro}

Medical image segmentation finds numerous application in clinical and translational imaging, including diagnosis, disease progression, treatment planning, and surgical assistance. Medical image segmentation frequently serves as the penultimate process in computer-aided diagnostic pipelines, particularly, in integrative multi-omics workflows \cite{Gillies2016Radiomics}.  
The gold standard approach to medical image segmentation is manual segmentation by clinical experts. Manual segmentation is time intensive, suffers from inter-observer and intra-observer variability, and poorly scalable to population studies involving large datasets. Deep segmentation networks (DSN) have emerged as an attractive substitute for manual segmentation where over-parameterized neural networks are trained on densely annotated medical scans in a fully-supervised fashion \cite{Azad2024UNetReview}. DSN can be classified into three broad categories: (a) convolutional neural networks (CNN) based architectures \cite{ibtehaz2020multiresunet, isensee2021nnu, ronneberger2015u}, (b) transformer based architectures \cite{azad2023daeformerdualattentionguidedefficient, hasan2025waveformer3dtransformerwaveletdriven, wu2024medsegdiff}, and (c) hybrid architectures \cite{wang2022smeswin, 54chen2021transunet, cao2022swin, wang2022uctransnet}.  

CNN-based DSN are (relatively) lightweight due to parameter sharing. The localized convolution-deconvolution operations, however, limit their receptive field, yielding suboptimal segmentation performance \cite{Azad2024UNetReview}. CNN-based models also exhibit poor generalizability, reporting substantial drop in performance on out-of-domain (OOD) datasets \cite{zhang2023understandingtricksdeeplearning}. Transformer-based architectures employ global self-attention to capture long-range (global) contextual information for enhanced segmentation performance \cite{shamshad2022transformersmedicalimagingsurvey}. Transformer models have high model complexity, require a large memory footprint, and focus on global contextual information, thereby neglecting spatial details at a local (patch) level \cite{Pang2024SlimUNETR}. Due to large number of trainable parameters, transformer models are prone to overfitting on small datasets, which compromises generalizability on OOD datasets. Hybrid architectures merge the strengths of CNN and transformer models, integrating local semantic information from convolution operations with global semantic information derived from self-attention modules \cite{wang2022uctransnet, wang2022smeswin, 54chen2021transunet, cao2022swin}. Hybrid architectures involve a tradeoff between accuracy and model complexity, with more sophisticated models offering higher performance but requiring greater computational resources. 

Recently proposed state-of-the-art DSN have high computational complexity and, therefore, require GPUs with substantial computing power (see Table \textbf{\ref{tab:compute}}). Even the lighweight architectures proposed for medical image segmentation are compute-intensive (Table \textbf{\ref{tab:compute}}). A notable exception is MA-TransformerV2 \cite{Wang2025Lightweight} which was trained on RTX 2080Ti GPU (11 GB, batch size = 2). However, training a DSN with a small batch size, on large datasets, significantly increases computation time and makes the training process unstable \cite{shen2023efficienttraininglargescaledeep}.

\begin{table}[htbp]
    \centering
    \small
    \scalebox{0.90}{
    \begin{tabular}{c|c|c}
        \hline
        \textbf{Model} & \textbf{GPU (VRAM)} & \textbf{Batch Size} \\
        \hline
        Swin-UNet        & V100 (32 GB)       & 24 \\
        UNETR++ \cite{shaker2024unetrdelvingefficientaccurate}         & A100 (40 GB)       & 4 \\
        UCTransNet       & A48 (48 GB)        & 4 \\
        Swin-UMamba \cite{liu2024swinumambamambabasedunetimagenetbased}     & A100 (40 GB)       & 1 \\
        SegMamba-V2      & A100 (40 GB)       & 2 \\

        U-Mamba \cite{ma2024umambaenhancinglongrangedependency}        & A100 (40 GB)       & 32 \\
        \hline
        MedSegDiff-V2    & A100 (40 GB)       & 32 \\
        SD-Seg           & V100 (16 GB)       & 4 \\
        GSS              & RTX A6000 (48 GB)  & 32 \\
        \hline
        MLRU++ \cite{yadav2025mlrumultiscalelightweightresidual}          & V100 (32 GB)       & 2 \\
        Slim UNETR       & V100 (32 GB)       & 16 \\
        GMS              & A100 (40 GB)       & 8 \\
        MA-TransformerV2 & RTX 2080Ti (11 GB) & 2 \\
        \hline
    \end{tabular}
    }
    \caption{Compute infrastructure of modern DSN methods.}
    \label{tab:compute}
\end{table}

For a more equitable deployment of AI tools in hospitals and healthcare facilities, we need DSN that offer high performance and can be trained on cost-effective GPUs with limited memory and large batch sizes. In this work, we propose Wave-GMS, a lightweight and efficient multi-scale generative model for medical image segmentation. It uses a trainable encoder to create high-quality latent representation from a multi-resolution decomposition of input image. The model leverages a compressed version of SD-VAE \cite{rombach2022high}, Tiny-VAE \cite{taesd2025}, to generate latent representations of input image and segmentation mask. A Latent Mapping Model (LMM) \cite{huo2024generative} learns the mapping from multi-resolution latent representation of input image to the corresponding segmentation mask representation. The predicted segmentation mask is decoded using Tiny-VAE’s pretrained decoder. Multi-resolution latents are aligned with Tiny-VAE's latents to improve cross-VAE compatibility. 

Wave-GMS has a substantially smaller number of trainable parameters, does not require loading memory-intensive pretrained vision foundation models, and supports training with large batch sizes on GPUs with limited memory. We conducted extensive experiments on four publicly available datasets (BUS, BUSI, Kvasir-Instrument, and HAM10000), demonstrating that Wave-GMS achieves state-of-the-art segmentation performance with superior cross-domain generalizability, while requiring only $\sim$2.6M trainable parameters.

\vspace{-0.3cm}
\section{Method}
\label{sec:method}
\vspace{-0.2cm}

\begin{figure*}[t]
    \centering
    \includegraphics[width=0.7\textwidth]{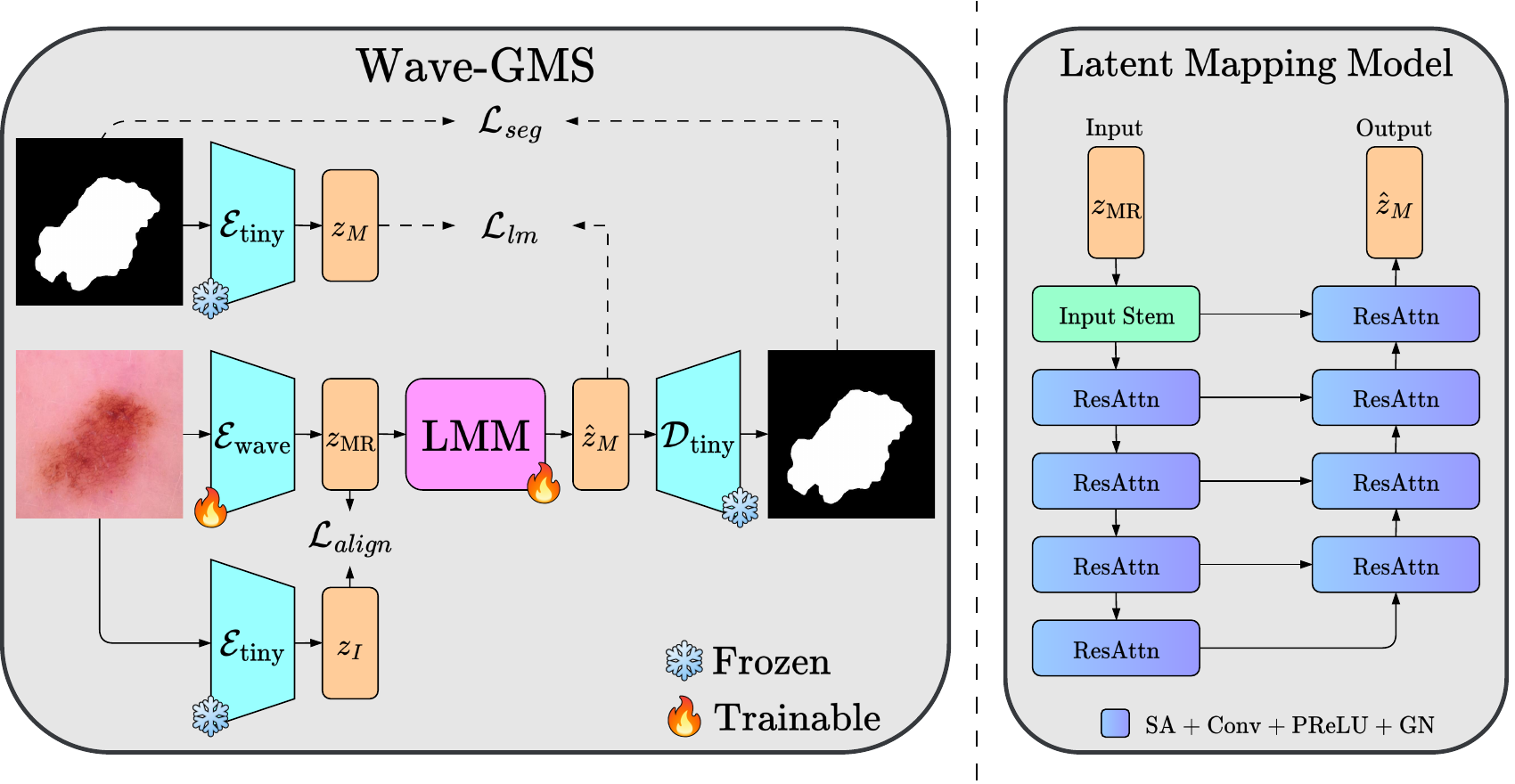}
            \caption{(\textit{Left}) Wave-GMS - A lightweight multi-scale generative model for medical image segmentation. (\textit{Right}) A latent mapping model (LMM) learns the transformation from the multi-scale latent space to the segmentation mask embedding space \cite{huo2024generative}.}
            \label{fig:Wave-GMS_pipeline} 
\end{figure*}

\textbf{Notation.} Grayscale images are denoted by $I \in \mathbb{R}^{H \times W}$ and RGB images are denoted by $I \in \mathbb{R}^{3 \times W \times H}$. Likewise, a binary segmentation mask is denoted by $M \in \mathbb{R}^{H \times W}$ and, if broadcasted to three channels (to match the dimension of an RGB image), is denoted by $M \in \mathbb{R}^{3 \times W \times H}$. $X_\text{MR} \in \mathbb{R}^{B \times 12 \times H/2^L \times W/2^L}$ denotes a multi-resolution decomposition of image $I$, where $L$ denotes the highest wavelet decomposition level. The output of the multi-resolution encoder, $\mathcal{E}_{\text{wave}}$, is denoted by $z_\text{MR} \in \mathbb{R}^{B \times 4 \times H/8 \times W/8}$. The output of the Tiny-VAE encoder, $\mathcal{E}_{\text{tiny}}$, is denoted by $z_I \in \mathbb{R}^{B \times 4 \times H/8 \times W/8}$ for the input image, and $z_M \in \mathbb{R}^{B \times 4 \times H/8 \times W/8}$ for the corresponding segmentation mask. The transformed segmentation mask representation (i.e., the output of LMM) is denoted by $\hat{z}_M \in \mathbb{R}^{B \times 4 \times H/8 \times W/8}$. $B$ denotes the batch size. The predicted segmentation mask in the image space (i.e., output of the Tiny-VAE decoder $\mathcal{D}_{\text{tiny}}$) is denoted by $\hat{M}$.

\subsection{Architecture Overview} The proposed Wave-GMS pipeline is illustrated in Figure \textbf{\ref{fig:Wave-GMS_pipeline}}. A trainable encoder, $\mathcal{E}_{\text{wave}}$, generates high-quality latent representations from a multi-resolution decomposition of input images~\cite{sadat2025litevaelightweightefficientvariational}: $z_\text{MR} = \mathcal{E}_{\text{wave}}(X_\text{MR})$. A highly compressed (distilled) version of the pretrained SD-VAE~\cite{rombach2022high}
---called Tiny-VAE~\cite{taesd2025}---generates latent representations of input images and corresponding segmentation masks: $z_I = \mathcal{E}_{\text{tiny}}(I)$ and $z_M = \mathcal{E}_{\text{tiny}}(M)$. A Latent Mapping Model (LMM) \cite{huo2024generative} learns a mapping from the multi-resolution latent representation of input images to the corresponding segmentation mask representation: $g_\theta^{\text{LMM}}: z_\text{MR} \rightarrow z_M$. A forward-pass through LMM generates the transformed segmentation mask representation: $\hat{z}_M = g_\theta^{\text{LMM}}(z_\text{MR})$. A predicted segmentation mask, $\hat{M}$, is obtained by decoding the transformed segmentation mask representation using the pretrained decoder of Tiny-VAE: $\hat{M} = \mathcal{D}_{\text{tiny}}(\hat{z}_M)$. The multi-resolution latents from the trainable encoder ($z_\text{MR}$) are aligned with the latent representation from the pretrained encoder of Tiny-VAE ($z_I$) to enhance cross-VAE compatibility. It must be noted that the Tiny-VAE (encoder and decoder) are kept frozen throughout the training routine. Only the lightweight encoder ($\sim$1.03M parameters) and the lightweight LMM ($\sim$1.56M parameters) are trained, which keeps the total number of trainable parameters substantially small ($\sim$2.6M).

\subsection{Multi-Resolution Encoder} 
Our multi-resolution encoder is inspired by~\cite{sadat2025litevaelightweightefficientvariational}. Each image, $I$, is processed using a multi-level 2D Discrete Haar Wavelet Transform (DWT) to obtain a multi-resolution decomposition: 
$$X_\text{MR}^l = [X_{LL}^l \Vert X_{LH}^l \Vert X_{HL}^l \Vert X_{HH}^l]$$ 
where $l$ denotes the wavelet decomposition level and $\Vert$ denotes the concatenation operator along the channel dimension. $X_\text{MR}^l$ has 12 channels: $3$ RGB channels $\times$ $4$ subband images. We use three decomposition levels to obtain an $8\times$ downsampling factor, i.e., $l \in \{1,2,3\}$. For each wavelet decomposition level, a feature-extraction module, $\phi_l$, computes a multi-scale set of feature maps: 
$$F_l = \phi_l(X_\text{MR}^l) = \phi_l([X_{LL}^l \Vert X_{LH}^l \Vert X_{HL}^l \Vert X_{HH}^l])$$ 
Since the resolution of feature maps, at distinct decomposition levels, differ by a dyadic factor of 2, we downsample the feature maps before concatenation: 
$$F = [\downarrow\downarrow(F_1) \Vert \downarrow(F_2) \Vert F_3]$$
where $\downarrow(\cdot)$ denotes downsampling by a factor of 2. Subsequently, a feature aggregation module, $\mathcal{A}$, combines the multi-scale feature maps from each decomposition level to yield the multi-resolution latent representation of the input image: 
$$z_{\text{MR}} = \mathcal{A}(F)$$ 
We employ a U-Net based architecture for the three feature extraction modules $\{\phi_l\}_{l=1}^3$ and the feature aggregation module $\mathcal{A}$, without spatial downsampling and upsampling layers~\cite{sadat2025litevaelightweightefficientvariational}. 
\begin{table*}[htbp]
\centering
\large
\resizebox{\textwidth}{!}{
\begin{tabular}{c|c|c|ccc|ccc|ccc|ccc}
\hline
\multirow{2}*{\textbf{Type}} & \multirow{2}*{\textbf{Model}} & \multicolumn{1}{c|}{\textbf{Trainable}} &
\multicolumn{3}{c|}{\textbf{BUS}} & \multicolumn{3}{c|}{\textbf{BUSI}} & \multicolumn{3}{c|}{\textbf{HAM10000}} & \multicolumn{3}{c}{\textbf{Kvasir-Instrument}} \\
\cline{4-15}
 &  & \textbf{Params (M)} & \textbf{DSC}$\uparrow$ & \textbf{IoU}$\uparrow$ & \textbf{HD95}$\downarrow$
                          & \textbf{DSC}$\uparrow$ & \textbf{IoU}$\uparrow$ & \textbf{HD95}$\downarrow$
                          & \textbf{DSC}$\uparrow$ & \textbf{IoU}$\uparrow$ & \textbf{HD95}$\downarrow$
                          & \textbf{DSC}$\uparrow$ & \textbf{IoU}$\uparrow$ & \textbf{HD95}$\downarrow$ \\
\hline
CNN & UNet \cite{ronneberger2015u}                 & 14.0  & 81.50 & 70.77 & 17.68 & 72.27 & 63.00 & 35.42 & 92.24 & 86.93 & 13.74 & 93.82 & 89.23 & 8.71 \\
CNN & MultiResUNet \cite{ibtehaz2020multiresunet}  & 7.3   & 80.41 & 70.33 & 19.22 & 72.43 & 62.59 & 34.19 & 92.74 & 87.60 & 13.02 & 92.31 & 87.03 & 9.49 \\
CNN & ACC-UNet \cite{ibtehaz2023acc}               & 16.8  & 83.40 & 73.51 & 16.49 & 77.19 & 68.51 & 25.49 & 93.20 & 88.44 & 10.83 & 93.91 & 89.73 & 8.74 \\
CNN & nnUNet \cite{isensee2021nnu}                 & 20.6  & 85.71 & 78.68 & 11.43 & 79.45 & 70.99 & 22.13 & 93.83 & 89.32 & 9.43 & 93.95 & \textbf{90.20} & 8.51 \\
CNN & EGE-UNet \cite{ruan2023ege}                  & 0.05  & 72.79 & 61.96 & 27.73 & 75.17 & 60.23 & 29.51 & 93.90 & 88.50 & 10.01 & 92.65 & 86.30 & 9.04 \\
\hline
Transformer & SwinUNet \cite{cao2022swin}          & 27.2  & 80.37 & 69.75 & 20.49 & 76.06 & 66.10 & 28.69 & 93.51 & 88.68 & 10.46 & 92.02 & 85.83 & 9.15 \\
Transformer & SME-SwinUNet \cite{wang2022smeswin}  & 169.8 & 78.87 & 67.13 & 22.19 & 73.93 & 62.70 & 30.45 & 92.71 & 87.21 & 12.53 & 93.32 & 88.27 & 8.91 \\
Transformer & UCTransNet \cite{wang2022uctransnet} & 66.4  & 83.44 & 73.74 & 16.33 & 76.55 & 67.50 & 25.46 & 93.45 & 88.73 & 10.91 & 93.27 & 88.48 & 8.84 \\
\hline
Generative & MedSegDiff-V2 \cite{wu2024medsegdiff} & 129.4 & 83.23 & 74.36 & 17.02 & 71.32 & 62.73 & 38.47 & 92.28 & 87.02 & 13.02 & 92.29 & 87.21 & 9.06 \\
Generative & SDSeg \cite{lin2024stable}            & 329.0 & 82.47 & 73.45 & 20.53 & 72.76 & 63.52 & 36.79 & 92.54 & 87.53 & 12.29 & 91.23 & 86.54 & 9.38 \\
Generative & GSS \cite{chen2023generative}         & 49.8  & 84.86 & 77.58 & 22.42 & 79.56 & 71.22 & 28.20 & 92.92 & 87.98 & 11.29 & 93.66 & 89.15 & \underline{7.25} \\
Generative & GMS \cite{huo2024generative}          & 1.56  & \underline{88.42} & \underline{80.56} & \underline{6.79} & \underline{81.43} & \underline{72.58} & \underline{19.50} & \textbf{94.11} & \textbf{89.68} & \underline{9.32} & \textbf{94.24} & \underline{90.02} & \textbf{7.03} \\
\hline
Generative & Wave-GMS (ours)  & 2.60  & \textbf{90.14} & \textbf{82.62} & \textbf{5.36} & \textbf{82.31} & \textbf{73.42} & \textbf{18.46} & \underline{93.93} & \underline{89.37} & \textbf{9.25} & \underline{94.00} & 89.40 & 9.24 \\
\hline
\end{tabular}
}
\caption{Quantitative segmentation performance on four datasets. The best and second-best performances are \textbf{bold}
and \underline{underlined}, respectively.}
\label{tab:all}
\end{table*}

\subsection{Latent Mapping Model (LMM)} 
The trainable LMM, $g_\theta^{\text{LMM}}$, is inspired by~\cite{huo2024generative}. LMM is an encoder-decoder (hybrid) architecture without upsampling and downsampling operations (Figure \textbf{\ref{fig:Wave-GMS_pipeline}}). The latent space resolution is preserved at $H/8 \times W/8$ with only the channel dimension modulated from 32 channels to 128 channels across four layers. An input stem processes $z_\text{MR}$ via a convolutional block to generate a feature vector with 32 channels. It is followed by four encoder and four decoder ResAttn blocks (Figure \textbf{\ref{fig:Wave-GMS_pipeline}}). Each block consists of a residual unit and a spatial self-attention layer. Skip connections between convolutional layers mitigate vanishing gradients and preserve semantically relevant features. LMM also includes deep supervision in the four decoder layers to enhance feature learning and regularize model training. Deep supervision was not applied during inference; the predicted segmentation mask was obtained from the last layer of the decoder.

\subsection{Training Loss Function} 
Wave-GMS employed the following loss function:
\begin{equation*}
    \mathcal{L}_{\text{total}} = \mathcal{L}_{\text{seg}} + \mathcal{L}_{\text{lm}} + \mathcal{L}_{\text{align}},
\end{equation*}
where $\mathcal{L}_{\text{seg}}$ is the soft-dice loss between the predicted segmentation mask, $\hat{M}$, and the ground-truth segmentation mask $M$. $\mathcal{L}_{\text{seg}}$ includes four soft-dice loss terms obtained from four (intermediate) decoder outputs (deep supervision). $\mathcal{L}_{\text{lm}}$ is a (deep supervision) $\ell_2$ reconstruction loss enforcing the predicted segmentation mask representation, $\hat{z}_M$, to match the latent representation of the ground-truth segmentation mask $z_M$. $\mathcal{L}_{\text{align}}$ promotes alignment between the multi-resolution latent space and the Tiny-VAE embedding space to enhance cross-VAE compatibility \cite{xu2025exploringrepresentationalignedlatentspace}:
\begin{equation*}
    \mathcal{L}_{\text{align}} = 0.9 \left( 1 - \cos(z_\text{MR}, z_I) \right) + 0.1 \, \|z_\text{MR} - z_I\|_1.
\end{equation*}
\section{Experiments}
\label{sec:experiment}

\subsection{Datasets}
We evaluated the performance of Wave-GMS on four publicly available datasets: BUS, BUSI, Kvasir-Instrument, and HAM10000. The BUS~\cite{yap2017automated} and BUSI~\cite{al2020dataset} datasets consist of breast lesion ultrasound datasets. BUS includes 163 subjects (132 train, 31 test), while BUSI contains 647 subjects (517 train, 130 test). Kvasir-Instrument~\cite{jha2021kvasir} comprises endoscopic images of 590 subjects (472 train, 118 test). HAM10000~\cite{tschandl2018ham10000} contains dermatoscopic scans of 10,015 subjects (8,015 train, 2,000 test). All datasets include manually annotated segmentation masks of regions of interest, provided by clinical experts. Images and corresponding masks were resized to $224\times224$ pixels.

\vspace{-0.25cm}

\subsection{Implementation Details}
Wave-GMS was implemented in PyTorch and trained on an RTX 3060 GPU (12 GB). We used the AdamW optimizer with a cosine annealing scheduler (initial learning rate of $2\times10^{-3}$). All experiments used a batch size of 12, a random seed of 2333, and 1000 training epochs -- except for HAM10000, which was trained for 300 epochs. Data augmentation involved random flipping, random rotations, and color jittering in the HSV domain~\cite{huo2024generative}. Model selection was based on the best validation Dice score. Segmentation performance was evaluated using DSC, IoU, and HD95 metrics.

\vspace{-0.1cm}
\subsection{Quantitative Results}
We compared Wave-GMS with other state-of-the-art DSN including five CNN-based models, three hybrid transformer-based models, and four generative models (see Table \textbf{\ref{tab:all}}). Wave-GMS outperformed all competing algorithms across the four benchmark datasets, achieving the highest DSC and IoU, and the lowest HD95 scores in every case -- except on the Kvasir-Instrument dataset where its performance was on par with GMS. Notably, Wave-GMS achieved these results with only $\sim$2.6M trainable parameters, making it one of the most lightweight models in Table \textbf{\ref{tab:all}}.

While GMS may appear more efficient in terms of trainable parameters, it relies on a heavyweight pretrained SD-VAE, which loads entirely onto the GPU. The SD-VAE encoder contains $\sim$34.2M parameters and the decoder $\sim$49.5M parameters, significantly increasing memory consumption and reducing the feasible batch size when training on RTX 3060 GPU (12GB). In contrast, Wave-GMS uses a highly compact pretrained Tiny-VAE, with only $\sim$1.22M parameters each for its encoder and decoder ~\cite{taesd2025}, enabling efficient training even on resource-constrained hardware.

Among generative models, Wave-GMS outperformed large-scale models such as SDSeg ($\sim$329M parameters) and MedSegDiff-V2 ($\sim$129.4M parameters), despite having the fewest trainable parameters. Since Wave-GMS has significantly fewer trainable parameters, it reduces the risk of overfitting on the training dataset -- especially on small training datasets.

\vspace{-0.2cm}

\subsection{Domain Generalization}
Table \textbf{~\ref{tab:generalizability}} presents a comparison of DSN for cross-data generalizability between two breast ultrasound datasets (BUS and BUSI). The BUS dataset was acquired at the UDIAT, Sabadell, Spain~\cite{yap2017automated}, with a Siemens ACUSON Sequoia C512 system, and the BUSI dataset was acquired at the Baheya Hospital, Cairo, Egypt, with LOGIQ E9 systems~\cite{al2020dataset}. Although both use ultrasound imaging, they have distinct data distributions due to diverse acquisition protocols and post-processing routines. For the cross-data generalizability study, we used the BUS and BUSI datasets alternately as training and test sets.

\setcounter{table}{3}
\begin{table*}[htbp]
\centering
\scalebox{0.90}{%
\begin{tabular}{l|ccc|ccc|ccc}
\hline
\multirow{2}{*}{\textbf{Model}} &
\multicolumn{3}{c|}{\textbf{BUS}} &
\multicolumn{3}{c|}{\textbf{BUSI}} &
\multicolumn{3}{c}{\textbf{Kvasir-Instrument}} \\
\cline{2-10}
 & \textbf{DSC}$\uparrow$ & \textbf{IoU}$\uparrow$ & \textbf{HD95}$\downarrow$ &
   \textbf{DSC}$\uparrow$ & \textbf{IoU}$\uparrow$ & \textbf{HD95}$\downarrow$ &
   \textbf{DSC}$\uparrow$ & \textbf{IoU}$\uparrow$ & \textbf{HD95}$\downarrow$ \\
\hline
Tiny-VAE (model mismatch)        & 86.24 & 77.88 & 9.48 & 79.02 & 69.97 & 20.79 & 93.79 & 89.33 & \underline{9.37} \\
Tiny-VAE (trained)               & 89.38 & 81.20 & 6.03 & 81.05 & 72.15 & \underline{17.64} & 92.08 & 86.88 & 14.25 \\
Tiny-VAE + MultiRes SFT          & 89.95 & 82.08 & 6.28 & 80.98 & 72.26 & 18.61 & 93.11 & 88.65 & 10.00 \\
Wave-GMS (w/o alignment)         & 89.54 & 81.49 & 6.11 & \underline{82.24} & \underline{72.88} & \textbf{16.91} & \underline{93.92} & \underline{89.36} & 9.68 \\
Wave-GMS (\texttt{batch\_size = 2}) & 89.84 & 81.96 & \underline{5.52} & 80.32 & 71.07 & 20.97 & 92.93 & 87.99 & 12.23 \\
Wave-GMS (\texttt{batch\_size = 4}) & \underline{90.11} & \underline{82.38} & 6.24 & 79.12 & 70.21 & 22.35 & 92.00 & 86.75 & 10.67 \\
Wave-GMS (with alignment)        & \textbf{90.14} & \textbf{82.62} & \textbf{5.36} & \textbf{82.31} & \textbf{73.42} & 18.46 & \textbf{94.00} & \textbf{89.40} & \textbf{9.24} \\ 
\hline
\end{tabular}
}
\caption{Ablation study. The best and second-best performances are \textbf{bold}
and \underline{underlined}, respectively. Unless otherwise specified, the batch size is $12$.}
\label{tab:ablation}
\end{table*} 
\setcounter{table}{2}
\begin{table}[ht]
\centering
\scalebox{0.95}{
\small
\setlength{\tabcolsep}{4pt}
\begin{tabular}{l|cc|cc}
\hline
\multirow{2}*{\textbf{Model}} &
\multicolumn{2}{c|}{\textbf{BUSI to BUS}} &
\multicolumn{2}{c}{\textbf{BUS to BUSI}} \\
\cline{2-5} & \textbf{DSC}$\uparrow$ & \textbf{HD95}$\downarrow$ & \textbf{DSC}$\uparrow$ & \textbf{HD95}$\downarrow$ \\
\hline
UNet           & 62.99 & 47.26 & 53.83 & 96.81 \\
MultiResUNet   & 61.53 & 53.97 & 56.25 & 94.31 \\
ACC-UNet       & 64.60 & 42.87 & 47.80 & 135.24 \\
nnUNet         & 78.39 & 20.53 & 59.13 & 89.32 \\
EGE-UNet       & 69.04 & 34.63 & 54.46 & 105.23 \\
\hline
SwinUNet       & 78.38 & 21.94 & 57.47 & 91.63 \\
SME-SwinUNet   & 74.78 & 25.81 & 58.28 & 91.26 \\
UCTransNet     & 72.76 & 28.47 & 56.94 & 94.32 \\
\hline
MedSegDiff-V2  & 69.56 & 32.51 & 55.21 & 98.57 \\
SDSeg          & 74.03 & 26.32 & 57.03 & 94.61 \\
GSS            & 68.74 & 35.74 & 58.72 & 92.57 \\
GMS            & \underline{80.31} & \underline{18.55} & \underline{61.60} & \underline{85.25} \\
\hline
Wave-GMS (ours) & \textbf{82.10} & \textbf{15.35} & \textbf{66.75} & \textbf{32.57} \\
\hline
\end{tabular}
}
\caption{Quantitative performance for domain generalization segmentation study.
The best and second-best performances are \textbf{bold}
and \underline{underlined}, respectively.}
\label{tab:generalizability}
\end{table} 

The proposed approach, Wave-GMS, significantly outperformed all competing methods in both transfer-directions. For the BUSI-to-BUS domain-transfer study, Wave-GMS achieved the highest Dice score (82.1\%) and lowest HD95 (15.35), indicating strong segmentation accuracy and precise delineation of region-boundary. In the BUS-to-BUSI domain-transfer study, Wave-GMS again reported the highest Dice score (66.75\%) and lowest HD95 (32.57), demonstrating strong robustness across diverse data domains. Compared to strong baselines like nnUNet, SwinUNet, and MedSegDiff-V2, Wave-GMS shows consistent improvements, highlighting its effectiveness in generalizing to unseen data distributions.

The remarkable performance of Wave-GMS is attributed to the high-quality, multi-resolution latent representation of the input image obtained with a lightweight, efficient, and trainable multi-resolution encoder. This is further enhanced by latent-space alignment with rich, domain-agnostic representations extracted from a distilled version (Tiny-VAE) of a pretrained large vision foundation model (SD-VAE).

\subsection{Ablation Studies}
Table \textbf{~\ref{tab:ablation}} presents an ablation study evaluating the impact of different training and alignment strategies on segmentation performance across three datasets: BUS, BUSI, and Kvasir-Instrument. \underline{Tiny-VAE (model mismatch)} is a training-free baseline experiment where the pretrained Tiny-VAE model was integrated with the pretrained LMM for each dataset (using LMM weights shared by~\cite{huo2024generative}). \underline{Tiny-VAE (trained)} integrated the pretrained Tiny-VAE model with a trainable LMM for each dataset. \underline{Tiny-VAE + Multi-Res SFT} injected multiresolution information in LMM using the spatial feature transform~\cite{wang2024exploitingdiffusionpriorrealworld, li2024extremeimagecompressionlatent}. The multi-resolution Haar coefficients, representative of high-frequency information (9 channels in total), are fused into three-channel feature maps using Selective Kernel Feature Fusion \cite{zamir2020learningenrichedfeaturesreal} before being passed to SFT blocks. \underline{Wave-GMS (w/o alignment)} does not promote latent-space alignment for cross-VAE generalizability. 

Tiny-VAE (model mismatch) performed the worst because the pretrained LMM weights were only aligned with the latent representation of SD-VAE. Tiny-VAE (trained) significantly improved segmentation performance across the three datasets. Tiny-VAE + MultiRes SFT further enhanced segmentation performance. Wave-GMS (w/o alignment) matched or surpassed Multi-Res SFT in most performance metrics. Wave-GMS’s combination of multi-resolution encoding and latent-space alignment enhances segmentation accuracy and robustness across diverse medical imaging domains.
\vspace{-0.2cm}
\section{Conclusion}
\label{sec:conclusion}
We propose Wave-GMS, a lightweight multi-scale generative model for medical image segmentation. Wave-GMS incorporates a lightweight trainable multi-resolution encoder to learn semantically rich representation of input images and a pretrained (frozen) Tiny-VAE to generate latent representation of segmentation masks. A lightweight trainable Latent Mapping Model maps the multi-scale image representation to corresponding segmentation mask representations. The output of LMM is decoded via the pretrained Tiny-VAE decoder. Multi-resolution latents are also aligned with Tiny-VAE’s latents to improve cross-VAE compatibility. Wave-GMS has a substantially smaller number of trainable parameters ($\sim$2.6M), does not require loading memory-intensive pretrained vision foundation models, and supports training with large batch sizes on GPUs with limited memory. Wave-GMS achieves state-of-the-art segmentation performance with superior cross-domain generalizability.

\indent 

\textbf{Limitations and future work.} The proposed Wave-GMS framework is, currently, applicable to 2D medical image analysis. The pretrained Tiny-VAE foundation model is a distilled version of the SD-VAE foundation model which is trained on a large-scale 2D imaging datasets. Future work involves extending Wave-GMS to 3D medical image analysis and exploring the efficacy of novel foundation models.

\newpage
\bibliographystyle{IEEEbib}
\bibliography{refs}

\end{document}